\documentclass[aps,prb,twocolumn,superscriptaddress,showpacs]{revtex4-1}
\usepackage{hyperref}
\usepackage[dvips]{graphicx}
\usepackage{amssymb,amsmath,amsfonts}
\usepackage[T1]{fontenc}
\usepackage{float}
\usepackage{color,soul}
\usepackage{booktabs}
\usepackage{color}
\usepackage{changes}
\usepackage{tabularx}
\newcommand{\bra}[1]{\langle #1|}
\newcommand{\ket}[1]{|#1\rangle}
\newcommand{\unit}{$(\hbar/e)(\Omega\times $cm$)^{-1}$}

\begin{document}

\title{Ultrathin SnTe films as a route towards all-in-one spintronics devices}

\author{Jagoda S\l awi\'{n}ska}
 \affiliation{Department of Physics, University of North Texas, Denton, TX 76203, USA}

\author{Frank T. Cerasoli}
  \affiliation{Department of Physics, University of North Texas, Denton, TX 76203, USA}

\author{Priya Gopal}
  \affiliation{Department of Physics, University of North Texas, Denton, TX 76203, USA}

\author{Marcio Costa}
 \affiliation{Brazilian Nanotechnology National Laboratory, CNPEM, 13083-970 Campinas, Brazil}

\author{Stefano Curtarolo}
 \affiliation{Center for Autonomous Materials Design, Duke University, Durham, NC 27708, USA}
 \affiliation{Materials Science, Electrical Engineering, Physics and Chemistry, Duke University, Durham, NC 27708, USA}

\author{Marco \surname{Buongiorno Nardelli}}
 \affiliation{Department of Physics, University of North Texas, Denton, TX 76203, USA}
 \affiliation{Center for Autonomous Materials Design, Duke University, Durham, NC 27708, USA}


\begin{abstract}
Spin transistors based on a semiconducting channel attached to ferromagnetic electrodes suffer from fast spin decay and extremely low spin injection/detection efficiencies. Here, we propose an alternative all-in-one spin device whose operation principle relies on electric manipulation of the spin lifetime in two-dimensional (2D) SnTe, in which the sizable spin Hall effect eliminates the need for using ferromagnets. In particular, we explore the persistent spin texture (PST) intrinsically present in the ferroelectric phase which protects the spin from decoherence and supports extraordinarily long spin lifetime. Our first-principles calculations followed by symmetry arguments revealed that such a spin wave mode can be externally detuned by perpendicular electric field, leading to spin randomization and decrease in spin lifetime. We further extend our analysis to ultrathin SnTe films and confirm the emergence of PST as well as a moderate enhancement of intrinsic spin Hall conductivity. The recent room-temperature observation of the ferroelectric phase in 2D-SnTe suggests that novel all-electric spintronics devices are within reach.
\end{abstract}

\maketitle

The idea of using electron spins in transistors for information transfer and processing lies at the heart of research in the area of spintronics.\cite{spintronics} However, after two decades of efforts the pioneering concept of spin transistor proposed by Datta and Das still suffers from two major performance issues impeding its use in applications.\cite{das, das_transistor} First is the low efficiency of spin injection and detection through ferromagnets caused by the conductivity mismatch at the interface. Second is the two-faced nature of spin-orbit interaction; it enables spin manipulation along the channel but it is adverse for spin lifetime and leads to spin randomization in diffusive transport regime. Several approaches were proposed to overcome these obstacles, including all-electric spin Hall transistors without the ferromagnets\cite{jungwirth, all-in-one} or devices protected from spin decoherence by uniform spin configuration known as a persistent spin helix (PSH).\cite{schliemann, bernevig} However, these successful realizations rely on precisely controled structures, such as semiconductor quantum wells,\cite{wojtowicz, kohda, kohda_review} which usually limits the operating temperatures to few kelvin, preventing any practical use.

In parallel with the progress in spin transistors, several multifunctional materials have been recently designed or rediscovered; some of them reveal intriguing quantum phenomena intimately related to dimensionality, topology and symmetries. Group IV-VI monochalcogenides (MX, M=Ge, Sn ; X=S, Se, Te) are narrow gap semiconductors widely used in conventional devices where they serve as thermoelectrics, ferroelectrics, optical filters and detectors, photocatalysts or sensors. Remarkably, their intriguing spin-dependent electronic properties remained unexplored over decades; GeTe and SnTe have been just recently recognized as excellent candidates to use in spintronics. In particular, they represent a class of so-called ferroelectric Rashba semiconductors (FERSC) whereby the spin degree of freedom coupled to the ferroelectricity manifests in the electrically tunable Rashba spin texture of bulk electronic states.\cite{fersc, silvia, liebmann, nanoletters, fe-gete, plekhanov} Moreover, sizable spin Hall effect (SHE) has been proposed in both materials,\cite{haihang, ohya} which opens a perspective to integrate different functionalities and construct ferromagnet-free spin devices.

In this paper, we put forward the idea of all-in-one spin transistor based on two-dimensional (2D) SnTe. In such a device the spin injection and detection can be accomplished via direct and inverse spin Hall effects, while the on/off state is manipulated through the electric control of spin lifetime along the channel. Specifically, an atomic-thick SnTe in a structural form of phosphorene was suggested to host a persistent spin wave mode enforced by the crystal space group symmetry.\cite{hosik} Uniform spin polarization along out-of-plane direction is linked to the ferroelectricity, or more precisely to the in-plane polar displacement. As an intrinsic property of the material, it does not require any fine-tuning between spin-orbit parameters in order to support an exceptionally long spin lifetime.\cite{tsymbal_psh} Here, we propose that such a spin configuration can be externally detuned by perpendicular electric field breaking the crystal symmetries. The spins are then dephased by electron scattering, which enables the realization of switch-off mechanism. Importantly, the injected spins are by construction parallel to spin-orbit field, thus they will be transported without precession, making the device robust against switching between different momenta and sub-bands with opposite spin textures in the presence of doping.

Finally, the ferroelectric ground state has been recently observed in mono- and multilayer SnTe structures and surprisingly, it seems to persist in the latter at room temperature.\cite{science_snte} Our density functional theory (DFT) calculations confirmed that the SHE/PSH combination can be realized in specific ultrathin SnTe films. In the light of above, the all-in-one spin transistors are an attractive hypothesis, further discussed in terms of phase robustness and alternative realizations.

\begin{figure*}[ht!]
    \includegraphics[width=0.97\textwidth]{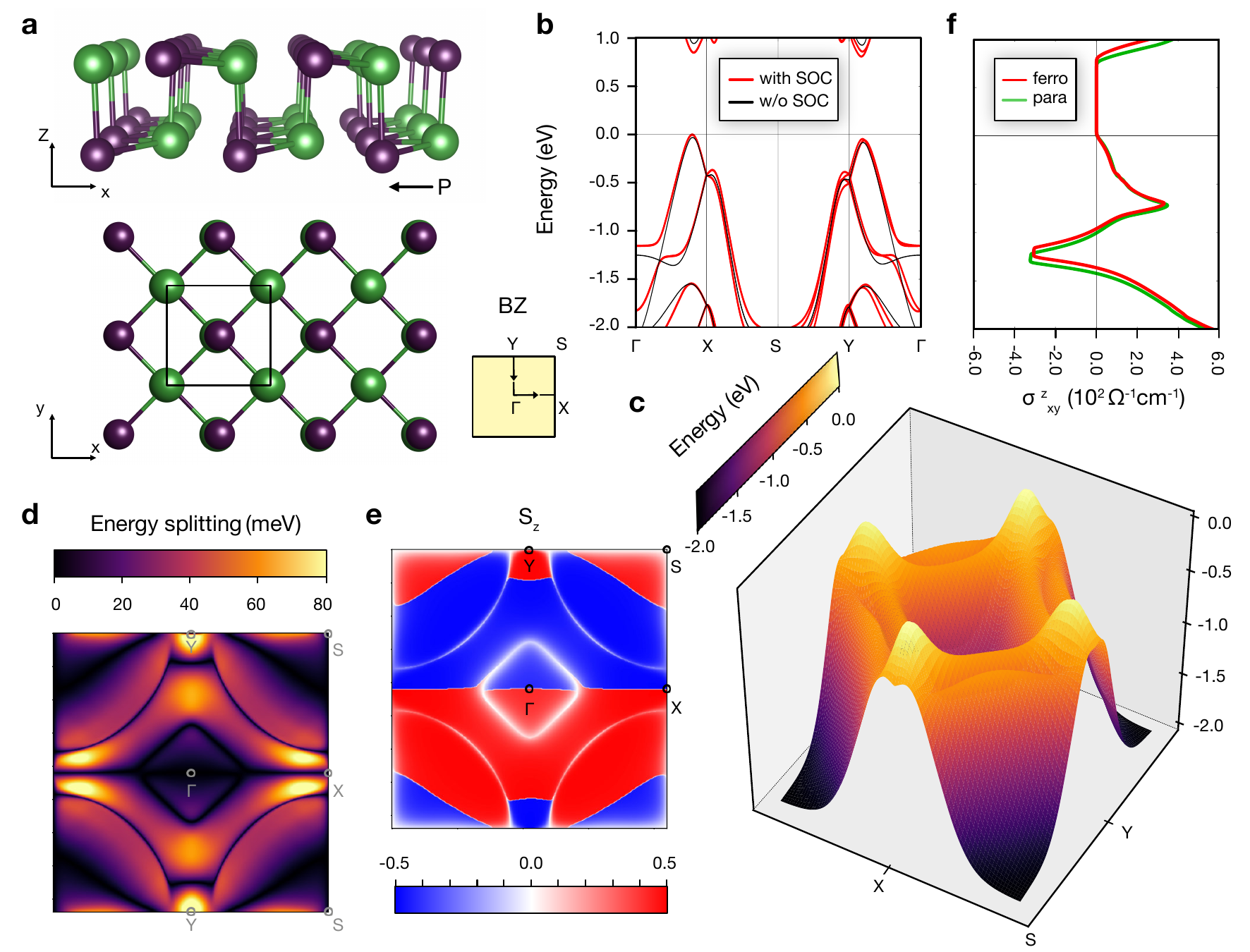}
    \caption{\label{fig1}
    Structural and electronic properties of SnTe monolayer. (a) Top and side view of the optimized structure. The Sn/Te atoms are displayed as green/purple spheres. Black rectangle denotes the unit cell and the inset shows the Brillouin zone. (b) Electronic structure calculated along high-symmetry lines marked in (a). Red/black lines represent the bands with/without including SOC. (c) Three-dimensional view of the topmost valence band over the entire BZ revealing its quasi-fourfold shape. The VBM is located close to $X$ along $\Gamma-X$ direction, but similar local maxima are present around $Y$. The energy splitting and the spin polarization of this band are presented in panels (d) and (e), respectively. The components $S_x$ and $S_y$ are negligible in this case, thus only $S_z$ component is shown. The labels marked in the maps denote high-symmetry points in the BZ, as defined in the inset in (a). (f) Spin Hall conductivity $\sigma^{z}_{xy}$  as a function of chemical potential calculated for the ferroelectric structure (red) shown in (a-b) and the corresponding paraelectric phase (green).
    }
\end{figure*}

\begin{figure*}[ht!]
    \includegraphics[width=0.97\textwidth]{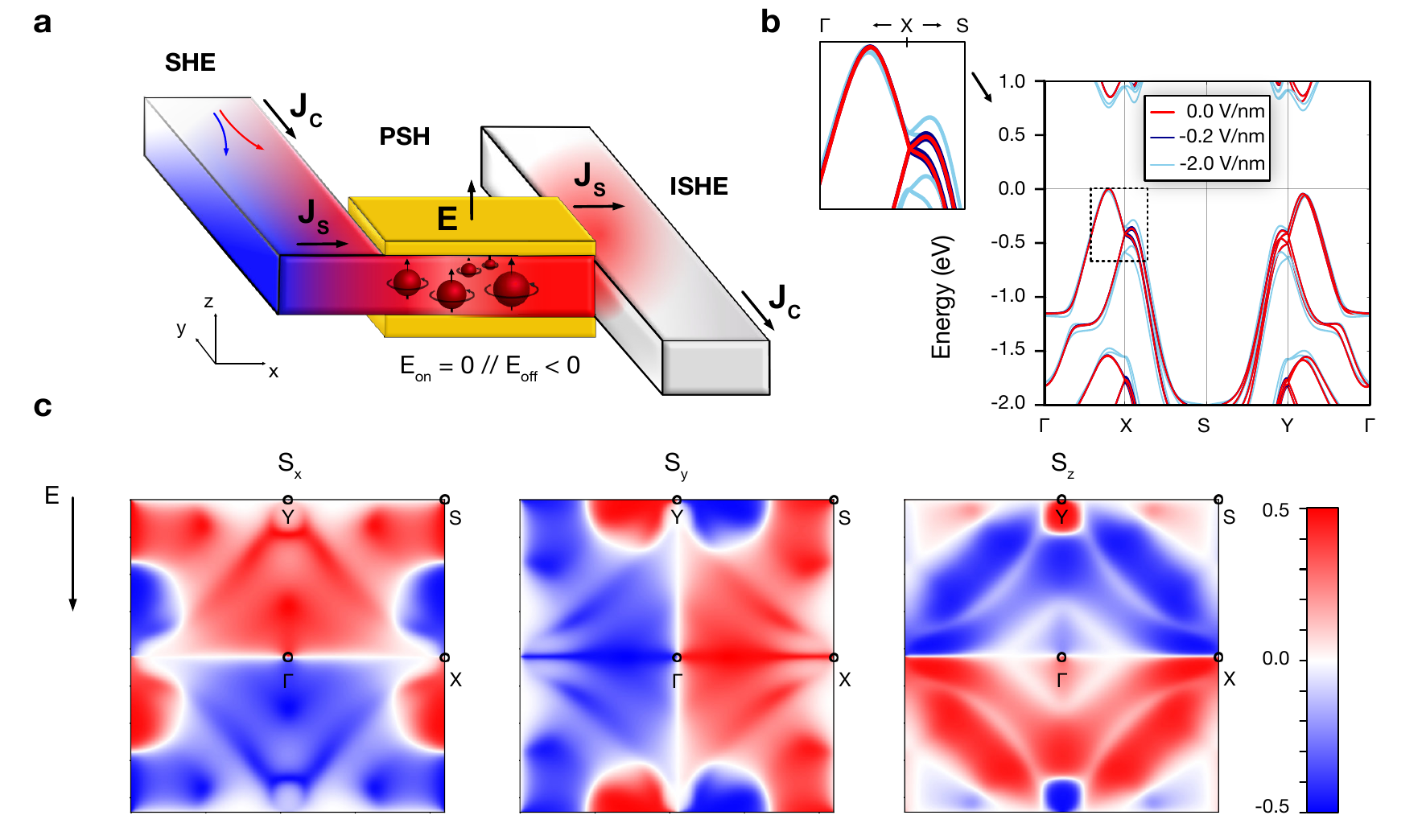}
    \caption{\label{fig2}
    Properties and applications of biased 1ML-SnTe. (a)
    Operation principle of all-in-one spin transistor based on 2D-SnTe. The spin injection is realized via SHE which induces the accumulation of spins polarized along $+z$. In the on state ($E=0$) the direction of the spin-orbit field is parallel to spin orientation, thus no spin precession occurs during the transport along the channel. The spin orientation is further detected based on the ISHE and measured as an induced voltage. In the off state ($E<0$) the PSH state is detuned by the electric field leading to spin decoherence. The ISHE is largely limited and the measured Hall voltage is negligible. The colors in the scheme denote the direction of spins, but do not reflect any numerical values. (b) Band structure of SnTe monolayer calculated for different values of electric field $E$. The inset shows the details around the VBM. (c) Spin polarization of the topmost valence band at $E = -2.0$ V/nm calculated over the entire BZ. We note that even close to the band maxima the components $S_{x}$ and $S_{y}$ are different from zero. The labels in the maps refer to high-symmetry points of the BZ.
    %Two gate electrodes are suggested to avoid unintended changes in the symmetry.
    }
\end{figure*}

\begin{figure*}[ht!]
    \includegraphics[width=0.97\textwidth]{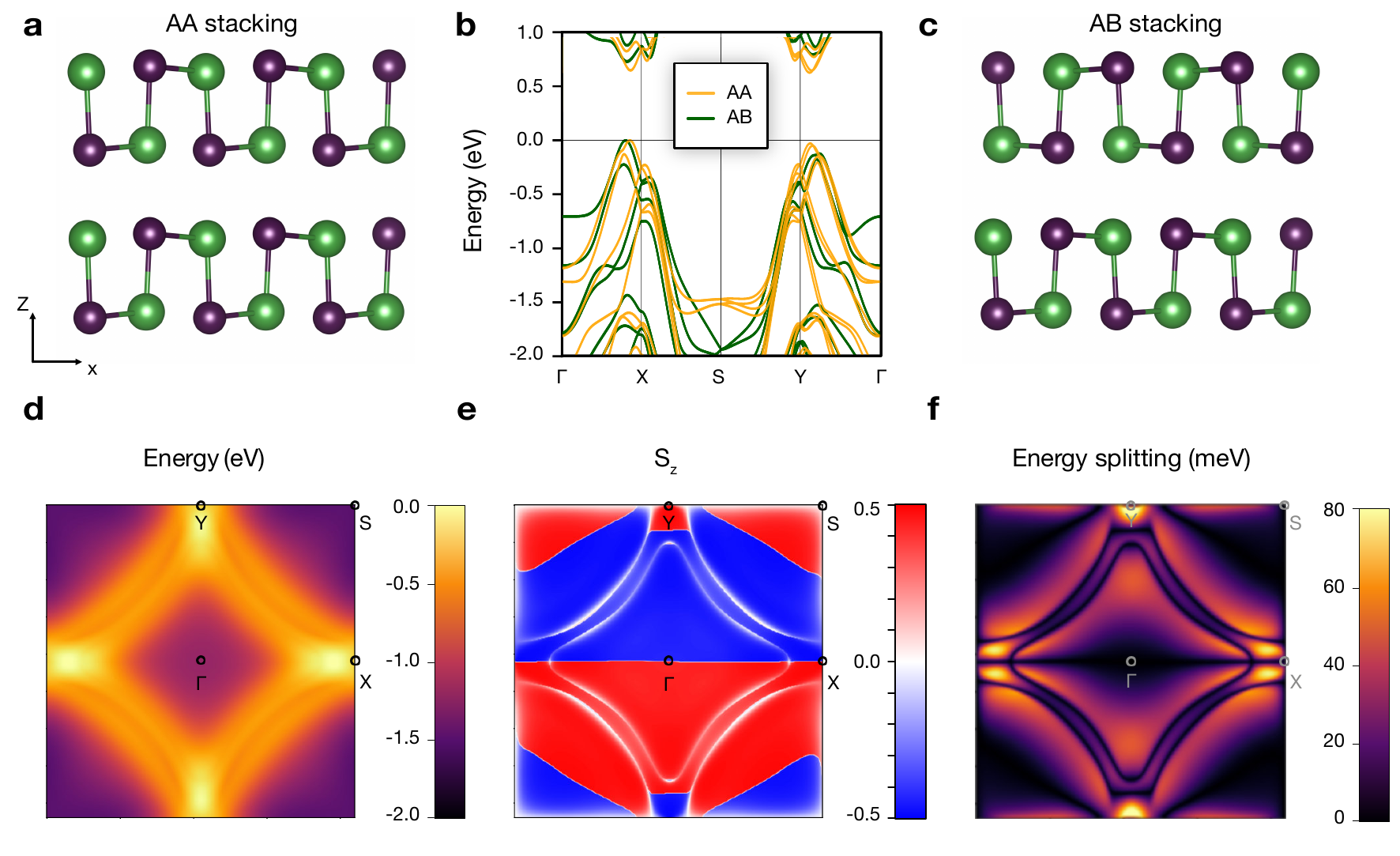}
    \caption{\label{fig3}
    Geometry and electronic structure of the 2ML-SnTe. (a) Optimized structure of the AA-stacked bilayer. (b) Band structure calculated for the AA and AB stacking configurations, represented by orange and dark green lines, respectively. (c) Relaxed geometry of the AB-stacked bilayer. (d) Momentum-resolved band topography, (e) the associated spin texture ($S_{z}$) and (f) the energy splitting of the topmost valence state calculated over the entire BZ. The negligible components $S_x$ and $S_y$ are omitted in (e).
 %Some of the curves do not reach the end of the energy scale due to the issues related to projectability of the bands at these energies.\cite{paoflow}
    }
\end{figure*}
\begin{figure*}[ht!]
    \includegraphics[width=0.97\textwidth]{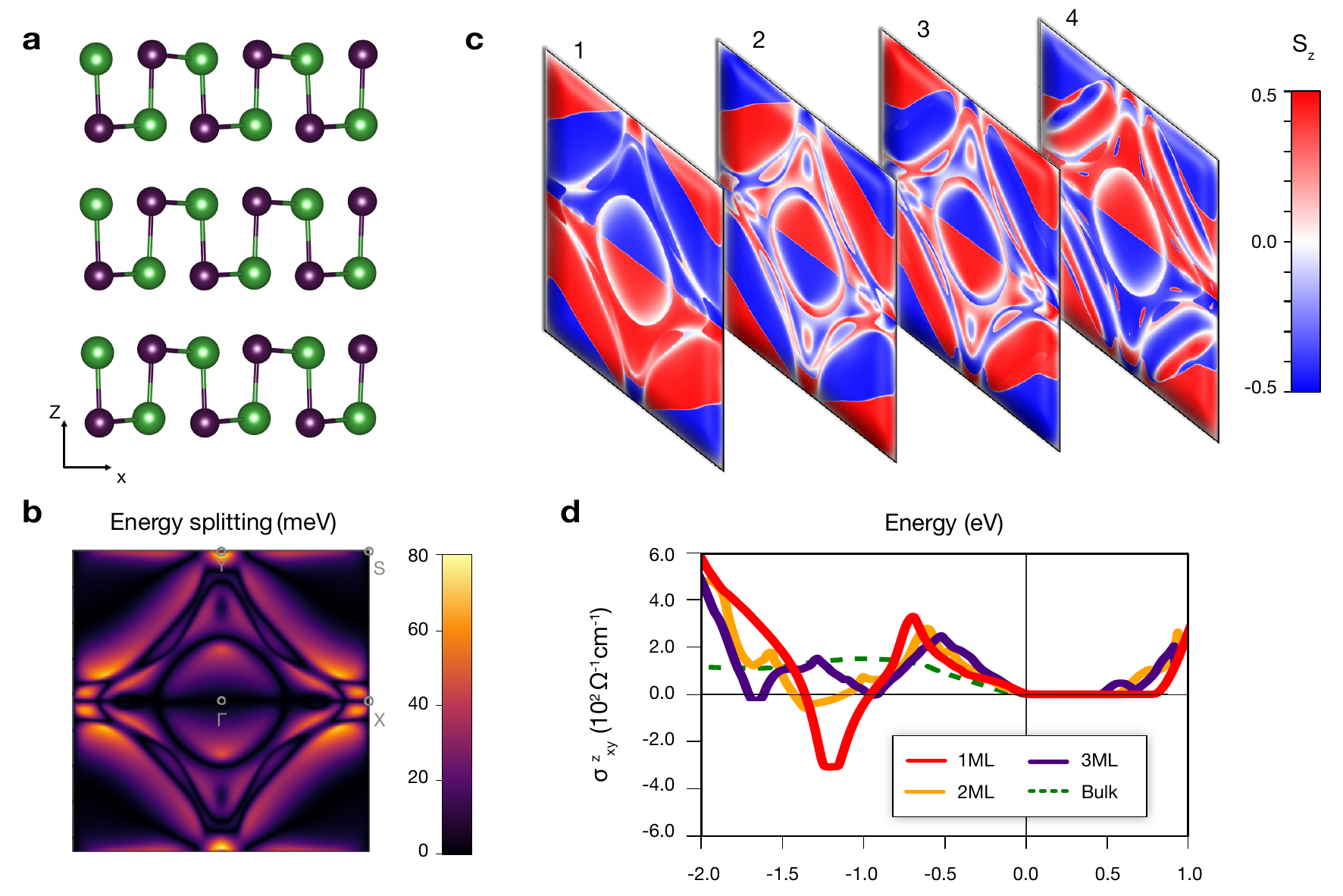}
    \caption{\label{fig4}
    Properties of multilayer SnTe. (a) Relaxed geometry of the 3ML-SnTe in the AA stacking configuration. (b) Momentum-resolved map of the spin-splitting in the topmost valence band calculated over the entire BZ for the structure shown in (a). (c) Spin textures of four topmost valence bands enumerated 1-4. The negligible in-plane components are omitted. (d) Spin Hall conductivity $\sigma^{z}_{xy}$  as a function of chemical potential calculated for different multilayers and bulk $\alpha$-SnTe in orthorhombic setting. Spin Hall conductivities of the multilayers are normalized by their effective volumes listed in Table I.
    }
\end{figure*}

Let us start with a brief overview of geometry, electronic properties and intrinsic spin Hall conductivity calculated for the SnTe monolayer (1ML-SnTe; in our notation, we refer to van der Waals monolayer equivalent to two atomic layers - 2AL\cite{parkin}). Figure \ref{fig1} (a)-(b) shows the orthorhombic lattice of the crystal. The ions are arranged in two buckled layers with distortions inducing the spontaneous polarization along $x$, while the tiny anisotropy between the lattice constants $a$ and $b$ emerges as a natural signature of the ferroelectric phase. Two-dimensional SnTe belongs to the space group no. 31 ($Pmn2_{1}$) invariant with respect to the following symmetry operations: (1) the identity operation $E$; (2) mirror reflection $M_{y}$ with respect to the $y=0$ plane, (3) glide reflection $\overline{M}_{z}$ consisting of mirror reflection $M_{z}$ about the $z=0$ plane followed by a fractional translation by a vector $\nu = (1/2a, 1/2b, 0)$ and (4) two-fold screw rotation $\overline{C}_{2x}$ combining two-fold rotation around the $x$-axis with the translation by $\nu$. Clearly, these symmetries suggest the intrinsic PSH in terms of the criteria formulated by Tao and Tsymbal.\cite{tsymbal_psh} Following their general arguments based on group theory, we can rationalize the fully (scalar) relativistic electronic structure of SnTe represented by red (black) lines in Figure \ref{fig1}  (b). In particular, the energy levels without the spin-orbit coupling (SOC) are fourfold degenerate, while the SOC splits them into doublets with eigenvalues of $\overline{M}_{z} = \pm1$ over the entire BZ except for the $\Gamma-X$ line. Furthermore, these doublets split into singlets with eigenvalues of spin operator $S_{z}$, indicating the persistent spin texture (PST) (anti-)aligned in the out-of plane direction.

Next, we will analyze the electronic states over the full Brillouin zone. We note that the unidirectional spin texture of SnTe was previously considered in terms of the effective $\vec{k}\cdot\vec{p}$ models which well describe the physics around the valence band maximum (VBM).\cite{hosik, snte_psh} However, in the presence of doping the regions beyond the VBM may also contribute to the spin transport. In Figure \ref{fig1} (c)-(e), we have plotted the topography of the topmost valence band along with the corresponding momentum-resolved map of the energy splitting and the spin texture. The three-dimensional view helps to localize the regions near the Fermi level which could potentially be reached at sufficient level of doping. Indeed, the maxima residing close to the $X/Y$ points are connected by saddle-like regions forming a shape similar to a four-point star. Importantly, from the map shown in Figure \ref{fig1} (d) we can conclude that the regions of the highest splitting well reflect the topography of the band; in the neighborhood of the VBM the values of splitting are as high as 80 meV, whereas around the saddle-like regions they achieve 50 meV. Such values are certainly sufficient to ensure proper functioning of devices at room temperature. Finally, the spin texture (Figure \ref{fig1} (e)) reveals only $S_z$ component consistent with the prediction based on the symmetries (the negligible components $S_{x}/S_{y}$ are omitted in the figure). The sign reversals present in (e), e.g. along the $\Gamma-Y$ line, seem to be associated to the swaping of sub-bands forming the Kramer's pairs.

Finally, we will turn our attention to the spin Hall conductivity reported in Figure \ref{fig1} (f). Due to the 2D nature of the structure (both charge and spin currents have to be in-plane), the only non-vanishing independent components of the SHC tensor are $\sigma_{xy}^{z}=-\sigma_{yx}^{z}$. As one could expect from the huge impact of the SOC on electronic states in the considered energy window (Figure \ref{fig1} (b)), the estimated magnitudes are relatively large, reaching almost 400 \unit\, at the first resonance peak. As derived in our previous analysis,\cite{2dmaterials} the $p$-type doping required to achieve such a value would have to be as large $10^{14}e/$cm$^2$ which is difficult to realize. In fact, experimental studies indicated a slight intrinsic $p$-type doping ($10^{11}e/$cm$^2$) in 2D-SnTe samples,\cite{science_snte, parkin} suggesting that SHC would be rather small. We note, however, that the spin Hall angles can still be sufficiently large because of the low charge conductivity; this may enable an efficient spin injection/detection in such a case. Importantly, we have not found any significant difference in SHC between the polar and centro-symmetric SnTe phases which ensures the realization of SHE even in the presence of doping.

The all-in-one spin transistor is designed based on the geometry illustrated in Figure \ref{fig2}(a). The spin injection is realized via the direct spin Hall effect in the left part of the device. The charge current along $y$ direction is converted into a spin current along $x$. The spins with out-of-plane polarization are then efficiently injected into a gate-controled region. The logic functionality is achieved by a purely electric manipulation of the anisotropy in spin lifetime determined by the presence or absence of the PSH mode, similarly to approaches employing semiconductor quantum wells.\cite{kohda_review} In the absence of electric field ($E=0$) the robust PSH maintains the spin polarization along $+z$ axis; we note that no spin precession occurs because the PST is always aligned or anti-aligned to the polarization of the spin current, whereby the sign depends on the electron's momentum and/or specific sub-band in a doped system. The spin current is then injected into the right part of the transistor and detected through the inverse spin Hall effect which induces the electric current along $y$ direction. On the contrary, setting $E\neq0$ perturbs the PSH mode leading to spin dephasing upon electron scattering. In this case the Hall voltage generated in the detecting region should be significantly reduced.
\vskip 1.0 em

Let us analyze in more detail the physical mechanisms that cause such a modulation of the PSH. Figure \ref{fig2} (b) compares the band structures of SnTe calculated at different values of $E$. It is clear that the external electric field lifts the valence band degeneracy along the $\Gamma-X-S$ line and further separates the energy of both pairs of doublets associated with the eigenvalues of $\overline{M}_{z}$ symmetry operator. While the electric potential along $z$ induces spatial anisotropy between the layers, the glide reflection will not be a valid transformation anymore, thus the observed change in the electronic structure along the high-symmetry lines is not surprising. Based on the general consideration performed by Tao and Tsymbal,\cite{tsymbal_psh} we note that for $E=0$ the $\overline{M}_{z}$ in spin space anticommutes with $\sigma_x$ and $\sigma_y$ which results in zero expectation values of both in-plane components of spin operator. Since such a condition cannot hold for $E\neq0$, we expect a severe change in the spin texture. Indeed, our DFT calculations confirmed the emergence of the in-plane spin components over the entire BZ, as shown in  Figure \ref{fig2} (c). Although around the VBM the $S_x/S_y$ are rather small, they can significantly detune the PST provided that a sufficiently strong electric field is applied.

We emphasize that the presence of PSH is strictly related to the space group (no. 31) of 2D-SnTe. This raises a question about the thickness limit for the proposed spin-transistor or more specifically, for how many layers the $Pmn2_{1}$ space group remains a valid description of the crystal. From the experimental side, ultrathin films form in a layered structural ($\gamma$) phase until around 3ML, while thicker samples may already consist of either mixed $\alpha$ and $\gamma$ phases at low temperatures, or $\beta$ and $\gamma$ phases at room temperature.\cite{parkin} This indicates an important limit for the proposed all-in-one spin device. In next paragraphs, we will report a more detailed analysis of 2ML and 3ML structures in different configurations. In particular, we will demonstrate that PSH/SHE combination can be still realized in such systems, which softens a strict criterion of a truly monolayer material.

Ultrathin SnTe films have been constructed by stacking single layers along vertical direction either in parallel (AA) or anti-parallel arrangement (AB),\cite{ferro_2017} corresponding to polar and antipolar configurations, respectively. We have further optimized the internal coordinates and lattice parameters imposing strict convergence criteria. Our calculations revealed a clear tendency to stabilize in AA configuration for each considered film thickness (see Table I, $\delta E$ denotes the energy difference between two stacking orders), but we will briefly compare the structural and electronic properties of polar and anti-polar structures by exploring the simplest example of 2ML-SnTe illustrated in Figure \ref{fig3} (a)-(c). At first glance, the geometries does not reveal any striking dissimilarity except for the direction of the polar displacement in the upper layer. However, the electronic structures represented by the orange and dark green lines corresponding to the AA and AB respectively, are dramatically different. While in the former the bands seem to consist of two slightly shifted/modified replica of the 1ML-SnTe valence states visible close to the Fermi level, in the latter the severe modification of the electronic structure suggests a more fundamental properties change. Indeed, a closer analysis of the crystal lattice reveals that the relaxed AA structure maintains the $Pmn2_{1}$ symmetry, while the AB configuration belongs to the space group no 62 ($Pnma$), which does not support the PST.\cite{tsymbal_psh} This observation is further confirmed by the complementary plots evaluated over the entire BZ; in the AA stacking, the topography of the topmost valence band (Figure \ref{fig3} (d)) is quite similar to its monolayer counterpart from Figure \ref{fig1} (c), and the spin texture contains only the $S_z$ component which confirms the emergence of the PSH mode. In addition, the momentum resolved map of the energy splitting calculated for this band suggests that in terms of spin-dependent electronic properties the relaxed AA configuration could be equally useful as the 1ML-SnTe. Finally, we note that the AB structure manifests an intricate spin texture (not shown), as expected from the difference in the space group symmetry.

Figure \ref{fig4} (a)-(c) reports the properties of 3ML-SnTe. The relaxed geometry reveals the slight differences in ferroelectric displacement between the subsequent atomic layers; they display an oscillating dependence with respect to the layer number, being the largest close to the surface (compare $d_\mathrm{out}$ and $d_\mathrm{in}$ parameters in Table I). This conclusion is consistent with the previous studies of ferroelectricity in SnTe films,\cite{silvia_snte, barriers} whereby the distinct surface and bulk like properties are explained in terms of different coordination of the inner and outer ions. Although the relaxation has slightly changed the ionic displacements and lattice constants , it has not altered the symmetry of the cell which still belongs to the $Pmn2_{1}$ space group. Also the electric polarization ($P_\mathrm{eff}$) is rather robust with increasing number of layers (see Table I). However, the analysis of the momentum-resolved map of the spin-splitting (Figure \ref{fig4} (b)) indicates that the PSH may become less stable for larger film thickness, especially at higher temperatures,  as the spin splittings do not exceed 50 meV and rapidly decrease away from the band maxima. We also emphasize a larger number of bands close to the Fermi level which, in principle, could contribute to spin transport even at moderate doping. As evident from Figure \ref{fig4} (c), the lower bands have a trend to lose the spin texture, which could also be detrimental for the PSH. Last, we note that the spin Hall conductivities (Figure \ref{fig4} (d)) hardly change with the number of layers which confirms that the PSH/SHE combination can be realized in SnTe films.

Finally, let us remark on the accuracy of our density functional theory approach. First, SnTe thin films are van der Waals (vdW) lone-pair ferroelectrics, whereby the hybridization interactions compete with the Pauli repulsion and strongly depend on the number of layers.\cite{silvia_snte, barriers} Although such systems are challenging to simulate, a reliable estimation of the geometry is a key ingredient of the analysis, as the emergence of spin texture is a direct consequence of the ferroelectric distortion. Our extensive tests of computational strategies confirmed the importance of vdW interactions; we have used the semi-empirical Tkatchenko-Scheffler approach which ensured a robust disortion as well as a stable convergence of the lattice in the region between the monolayer and the bulk. Notably, a similar method has been successfully used in a recent study of elemental Te, a peculiar vdW ferroelectric with in-plane polarization due to the interlayer interaction between the lone pairs.\cite{elemental_te} Second, one has to determine the favorable stacking of the multilayer. We have found the strong preference for polar (AA) structure (see Table I), which is in reasonable agreement with the previous theoretical calculations employing atomic orbitals basis.\cite{multilayers} Surprisingly, experimental results from the same study indicate that the anti-polar configurations are more stable which may be assigned to an interplay of several factors, such as details of the growth (bottom-up vs cutting from the bulk), coexistence of different phases ($\alpha$, $\beta$, $\gamma$),\cite{parkin, apl_materials} or the influence of the substrate.\cite{substrates} Last, we emphasize that the calculated intrinsic spin Hall conductivities are extremely sensitive to details of the electronic structure. In order to obtain valid predictions, we have used pseudo-hybrid Hubbard self-consistent approach ACBN0 which was previously demonstrated to provide an excellent accuracy in case of the bulk phase.\cite{haihang}

In summary, we have demonstrated that the all-in-one spin transistors employing the combined effect of PSH and SHE can be constructed based on the SnTe multilayers. Such devices could benefit from the long spin lifetime ensured by the PSH as well as the efficient spin/charge interconversion without ferromagnetic electrodes. However, one needs to have in mind several important conditions that have to be satisfied. First, a natural thickness limit for such a device would be of 3ML-SnTe ($\sim$20 \AA), as concluded from out first-principles calculations and earlier experimental results. Second, the emergence of PSH is strictly determined by the crystal space group; it is possible only for multilayers with a polar stacking order, which preserve $Pmn2_{1}$ symmetry. Third, the realistic predictions of a spin transistor performance should take into account the role of the substrate. In particular, the interface could change the favorable stacking of vdW layers yielding an anti-polar instead of the polar order. Moreover, the presence of the substrate may itself perturb the spin wave mode. In such a case, the spins would not be fully protected from relaxation and the disorder could cause a decrease in spin lifetime, as observed in similar two-dimensional devices.\cite{roche1, roche2} In order to maximally preserve the spatial symmetry, the system should preferably have a sandwich structure.

We believe that the presented results will stimulate a further search of low-dimensional structures with similar properties. We also note that, in general, the simultaneous use of PSH and SHE does not need to be limited to 2D ferroelectrics. There are several bulk materials with persistent spin texture (e.g. BiInO$3$, LiTeO$3$, CsBiNb$_2$O$_7$ or Bi$_{2}$WO$_{6}$);\cite{tsymbal_psh, csbinb2o7, djani} they may also reveal a strong spin Hall effect and serve in all-in-one spin devices with a different operation principle.

\begin{table}
\caption{Calculated parameters of SnTe thin films. Lattice constants $a$ and $b$, the effective thickness $d_\mathrm{eff}$ and the ferroelectric displacements of the innermost ($d_\mathrm{in}$) and outermost ($d_\mathrm{out}$) atomic layers are expressed in angstroms. The energy difference between $AA$ and $AB$ stacking configurations ($\delta E$) as well as the band gaps ($E_{g}$) are given in electronvolts. The effective polarization ($P_\mathrm{eff}$) normalized to the thickness $d_\mathrm{eff}$ is expressed in $\mu$C/cm$^2$.
%Bulk lattice constants are equal to 4.50 and 4.49 \AA.
}
\vskip 1.0 em
\begin{tabularx}{0.95\columnwidth}{XXXXXXXXX}
\hline
&  $a$ &   $b$    &$  d_\mathrm{eff}  $   &$  d_\mathrm{in}  $   &$  d_\mathrm{out}  $   &$  \delta E  $  &$  P_\mathrm{eff}  $  &$  E_\mathrm{g}  $ \\
\hline
1ML& 4.56  & 4.53  &  9.8  &  -    & 0.11  &   -    & 10.2  & 0.81\\
2ML& 4.53  & 4.50  & 16.4  & 0.10  & 0.13  & -0.17  & 13.0  & 0.63\\
3ML& 4.52  & 4.49  & 23.1  & 0.10  & 0.13  & -0.39  & 13.4  & 0.51\\
%Bulk&4.50 &4.49 &n/a   &n/a  &n/a   &n/a    &n/a  &x\\
\hline
\hline
\end{tabularx}
\end{table}
\vskip 1.0 em

\textbf{Methods.} Our calculations based on density functional theory (DFT) were performed using the \textsc{Quantum Espresso} package.\cite{qe,qe1} We treated the ion-electron interaction with the norm-conserving pseudopotentials from the pslibrary database\cite{pslibrary} and expanded the electron wave functions in a plane wave basis set with the cutoff of 150 Ry. The exchange and correlation interaction was taken into account within the generalized gradient approximation (GGA) parameterized by the Perdew, Burke, and Ernzerhof (PBE) functional.\cite{pbe} We modeled the SnTe multilayers within the slab approach minimizing the errors introduced by the periodic boundary conditions by the large vacuum region of at least 20 \AA\, and with dipole corrections added to the local potential. We fully relaxed the structures setting the convergence criteria for energy and forces to $10^{-7}$ Ry and $10^{-4}$ Ry/bohr, respectively. The Tkatchenko-Scheffler van der Waals corrections were included in order to assure the stability and reliable lattice parameters of weakly interacting layers.\cite{vdw-ts} The electronic structures were further corrected by using a novel pseudo-hybrid Hubbard self-consistent approach ACBN0,\cite{acbn0} with the calculated $U$ parameters equal to 0.17 and 2.15 eV for Sn and Te, respectively.
%The Brillouin zone integrations were performed using the Monkhorst-Pack scheme with $k$-points grids of up to $24\times24\times1$.
%The electric polarizations were calculated using the Berry phase method.\cite{berry_phase}
Although the SOC was included self-consistently in DFT calculations, the spin-orbit related quantities were evaluated as a post-processing step employing the tight-binding Hamiltonians; the latter were constructed from the projections of eigefunctions on pseudoatomic orbitals following the implementation in the \textsc{PAOFLOW} code.\cite{paoflow}  After interpolating the Hamiltonians to an ultra-dense $k$-points mesh of $140\times140\times1$, we calculated the spin polarization of each eigenstate $\psi(\vec{k})$ represented as $S(\vec{k}) = [S_{x}(\vec{k}), S_{y}(\vec{k}), S_{z}(\vec{k})]$, where $S_{n}(\vec{k}) = \bra{\psi(\vec{k})}\sigma_{n}\ket{\psi(\vec{k})}$ and $\sigma_{n}$ denote the Pauli matrices. Spin Hall conductivities were computed from the Kubo's formula following the details given elsewhere.\cite{2dmaterials} Finally, we modeled the influence of electric fields perpendicular to the layers by modifying the tight-binding Hamiltonians.
\newline
\newline
%\begin{acknowledgments}
\textbf{Acknowledgments.} The members of the AFLOW Consortium  (http://www.aflow.org) acknowledge the grant ONR-MURI N000141310635. The authors also acknowledge Duke University --- Center for Materials Genomics --- and the CRAY corporation for computational support. Finally, we are grateful to the High Performance Computing Center at the University of North Texas and the Texas Advanced Computing Center at the University of Texas, Austin.
%\end{acknowledgments}

\bibliography{manuscript}% Produces the bibliography via BibTeX.

\end{document}